\documentclass{iaus}
\usepackage{graphicx}

\title[short title of paper] 
{Magnetism in the nearby galaxy M\,33}

\author[Tabatabaei, Krause, Beck, Fletcher]   
{F.~S. Tabatabaei$^1$,
 M. Krause$^1$, R. Beck$^1$, \and A. Fletcher$^2$}

\affiliation{$^1$Max-Planck-Institut f\"ur Radioastronomie, Auf dem H\"ugel 69,
53121 Bonn, Germany \break email: tabataba@mpifr-bonn.mpg.de, mkrause@mpifr-bonn.mpg.de,
rbeck@mpifr-bonn.mpg.de\\[\affilskip]
$^2$School of Mathematics and Statistics, Newcastle University,
Newcastle upon Tyne,\break NE1 7RU, U.K., email:
andrew.fletcher@ncl.ac.uk}

\pubyear{2009}
\volume{259}  
\pagerange{100--100}
\date{"Dec. 01, 2008" and in revised form ??}
\jname{Cosmic Magnetic Fields: From Planets,
to Stars and Galaxies}
\editors{K.G. Strassmeier, A.G. Kosovichev \&
 J.E. Beckman, eds.}
\begin{document}

\maketitle

\begin{abstract}
Using high-resolution data of the linearly polarized intensity and
polarization angle at 3.6, 6.2, and 20\,cm together with a 3-D model
of the regular magnetic field, we study variations of the structure,
strength, and energy density of the magnetic field in the Scd galaxy
M\,33. The regular magnetic field consists of a horizontal component
(represented by an axisymmetric mode from 1 to 3\,kpc radius and a
superposition of axisymmetric and bisymmetric modes from 3 to 5\,kpc
radius) and a vertical component. However, the inferred `vertical
field' may be partly due to a galactic warp. We estimate the average
total and regular magnetic field strengths as $\simeq$ 6.4 and
2.5\,$\mu$G, respectively. Generation of interstellar magnetic
fields by turbulent gas motion in M\,33 is indicated as the
turbulent and magnetic energy densities are about equal.
\end{abstract}

\firstsection 
\section{Introduction}
M\,33, the nearest Scd galaxy at a distance of 840\,kpc, with its
large angular size and medium inclination, allows determination of
the magnetic field components both parallel and perpendicular to the
line of sight equally well. The RM studies of M\,33 by
\cite{Beck_79} and \cite{Buczilowski_etal_91} suggested a
bisymmetric regular magnetic field structure in the disk of M\,33.
However, due to the low-resolution (1.8\,kpc) and low-sensitivity of
their observations, these results were affected by high uncertainty
particularly in the southern half of M\,33. Our recent observations
of this galaxy provide high-resolution (0.7\,kpc) maps of total
power and linearly polarized intensity at 3.6\,cm, 6.2\,cm, and
20\,cm presented by \cite{Tabatabaei_2_07}. These data are ideal to
study the rotation measure (RM), the structure and strength of the
magnetic field, and depolarization effects in detail. By combining
an analysis of multi-wavelength polarization angles with modeling of
the wavelength-dependent depolarization, \cite{Fletcher_04} and
\cite{Berkhuijsen_97} derived the 3-D regular magnetic field
structures in M\,31 and M\,51, respectively. The high sensitivity of
our new observations allows a similar study for M\,33.


\section{Nonthermal degree of polarization}\label{sec:pol}

Using the polarized intensity maps of \cite{Tabatabaei_2_07} and the
nonthermal maps obtained by \cite{Tabatabaei_3_07}, we derived maps
of the nonthermal degree of polarization at different wavelengths.
Integrating the polarized and nonthermal intensity maps in the
galactic plane out to a galactocentric radius of R~$\leq$~7.5~kpc,
we obtained the flux densities of the nonthermal and linearly
polarized emission and the average nonthermal degrees of
polarization of 10.3\%\,$\pm$\,2.0\%, 11.3\%\,$\pm$\,1.9\%, and
6.6\%\,$\pm$\,0.6\% at 3.6\,cm, 6.2\,cm, and 20\,cm, respectively,
indicating considerable wavelength-dependent depolarization by
Faraday effects at 20\,cm.

\section{The regular magnetic field structure}

In order to identify the 3-D structure of the \emph{regular}
magnetic field $B_{\rm reg}$ we fit a parameterized model of $B_{\rm
reg}$ to the observed polarization angles at different wavelengths.
We find that the Fourier modes $m=0+z0+z1$ ($z0$ and $z1$ are the
first and second Fourier modes of the vertical field) in the 1--3
kpc ring and  $m=0+1+z1$ in the 3--5 kpc ring can best reproduce the
observed pattern of polarized intensity at 6.2\,cm (see
\cite{Tabatabaei_1_08} for details). The horizontal magnetic field
component follows an arm-like pattern with pitch angles smaller than
those of the optical arm segments, indicating that large-scale
gas-dynamical effects such as compression and shear are not solely
responsible for the spiral magnetic lines. The dominant axisymmetric
mode ($m=0$) in the disk in both rings indicates that galactic
dynamo action is present in M\,33. We also find that the fitted
`vertical field', in the outer ring, could be mainly due to the
severe warp of M\,33 and hence apparent. However, a real vertical
field of a broadly comparable strength to the disk field can exist
in the inner ring.

\section{Magnetic field strengths and energy densities}
\label{subsec:equipart}

The strengths of the total magnetic field $B_{\rm tot}$ and its
regular component $B_{\rm reg}$ can be found from the total
synchrotron intensity and its degree of linear polarization.
Assuming equipartition between the energy densities of the magnetic
field and cosmic rays leads to  $B_{\rm tot} =\, 6.4\,\pm
\,0.5\,\mu$G  and $B_{\rm reg} =\, 2.5\,\pm\,1.0\,\mu$G for the disk
of M\,33 ($R<$\,7.5\,kpc).

\begin{figure}
\begin{center}
\resizebox{6cm}{!}{\includegraphics*{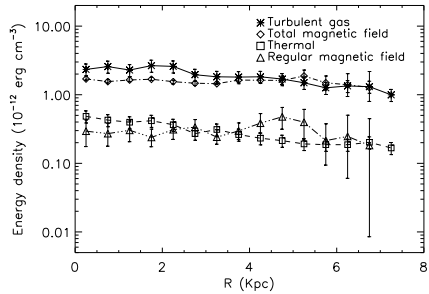}}
\caption{Energy densities and their variations with galactocentric
radius in M\,33.} \label{fig:energy}
\end{center}
\end{figure}

The energy densities of the equipartition magnetic fields in the
disk ($B_{\rm tot}^2 /8 \pi$ and $B_{\rm reg}^2/8 \pi$ for the total
and regular magnetic fields, respectively) are shown in
Fig.~\ref{fig:energy}. The energy densities of the magnetic field
and turbulence are about the same, confirming the theory of
generation of interstellar magnetic fields from turbulent gas
motion. Furthermore, it seems that the ISM in M\,33 can be
characterized by a low-$\beta$ plasma and is dominated by supersonic
turbulence, as the energy densities of the magnetic field and
turbulence are both much higher than the thermal energy density.

\end{document}